# Widely Tunable, Etch-free Multi-dielectric Fano Metasurfaces in the Visible

Md Rumon Miah[†] and Hilmi Volkan Demir*[,†,‡]

[†] Department of Electrical and Electronics Engineering, Department of Physics, UNAM − Institute of Materials Science and Nanotechnology and The National Nanotechnology Research Center, Bilkent University, Ankara 06800, Turkey.

[‡] Luminous! Center of Excellence for Semiconductor Lighting and Displays, School of Electrical and Electronic Engineering, Division of Physics and Applied Physics, School of Physical and Mathematical Sciences, School of Materials Science and Engineering, Nanyang Technological University, Singapore 639798, Singapore.

**ABSTRACT**

Fano resonances, characterized by a unique, sharp, and asymmetric spectral line shape arising from the interference between a narrowband discrete resonant state and a broadband continuum state, provide rich opportunities to control and manipulate light-matter interactions at the nanoscale. Here, we propose and demonstrate a nanophotonic architecture that integrates a multi-layered dielectric cavity into a one-dimensional grating to achieve an outstanding range of Fano resonance tunability by design, spanning 194 nm across the visible (from 500 to 694 nm), even with an ultra-thin layer of cavity (80 nm thick), offering the potential of highly compact photonic devices. The precise tailoring of the Fano spectral position through the adjustment of the cavity's structural parameters enables strongly localized field accumulation (at a maximum of 26 folds) within the cavity region. Additionally, the proposed architecture allows for bright and saturated structural colors in the visible, promising a route toward advanced digital display and printing technologies. Importantly, the experimental performance achieved with a simple, etch-free fabrication process results in a significant reduction in fabrication complexity and process-induced optical degradation. Our work showcases a robust platform for possible applications in Fano resonance-based low-loss nanolasers, on-chip photonic devices, optical communication components, and next-generation quantum technologies in the visible range.

## INTRODUCTION

Subwavelength resonant nanostructures specifically engineered to support sharp resonances make one of the fundamental groundworks of modern nanophotonic devices. Among them, Fano-type resonances are of critical importance and scientific interest[1,2,3,4]. The distinctive asymmetric spectral line shape of a Fano resonance emerges due to interference between a broadband continuum state and a narrowband discrete resonant state, enabling the control of light for enhanced emission as well as localization into engineered nanostructures[5,6,7,8,9]. Since its initial discovery in atomic systems[5,10], the Fano effect has been extensively explored in advanced nanophotonics, including plasmonic nanoparticles[11,12], dielectric nanostructures[13,14,15], and metasurfaces[16,17,18].

Dielectric nanostructures feature several key advantages compared to their plasmonic counterparts, most notably very low Ohmic losses, the distinct capability to support Mie-type magnetic resonances, and a relatively higher damage threshold[19,20]. These advantages make such dielectric nanostructures a superior platform for achieving high-quality resonances with enhanced spectral selectivity[21,22,23]. Remarkably, the low-loss dielectric nanostructures lead to experimental quality-factors approaching the million-scale ($10^6$)[24,25]. Fano resonance-based dielectric nanostructures possess a remarkable ability to produce sharp, asymmetric spectral resonances, intense light-matter interactions, and are highly sensitive to environmental changes[26,27,28]. Particularly, their capability to confine light energy in the cavity structures is an essential feature for the photoluminescence enhancement[29,30], lasing[31,32,33], and photonic nonlinear applications[34,35]. In addition, the dielectric Fano resonances facilitate a versatile platform for optical filtering[36], structural color generation[37,38], optical switching[39], high-performance sensing[18,28,40], modulation[41], and polarization control[42]. The growing interest in Fano resonance has been explored through subwavelength-scale gratings[27,43], nanoholes[29,44], nanodisks[15,30], and asymmetric structures[16,20,26]. Despite notable progress in dielectric Fano resonance nanostructures, prior studies have reported very narrow spectral tunability in the visible spectrum, where material selection opportunities are limited, often requiring a complex redesign process involving etching or pattern transfer and requiring strict limitations in fabrication tolerances[45,46,47,48].

Here, we propose and demonstrate both numerically and experimentally a nanophotonic architecture that integrates a multi-layered dielectric cavity with a dielectric grating layer to excite tunable Fano resonances across the visible spectrum. In the first approach, tunability was achieved by changing the refractive index, n, of the middle layer of the cavity design, while keeping the two outer layers identical (both made of $TiO_2$). The proof-of-concept demonstration was performed using three materials: polymethyl methacrylate (PMMA, n = 1.49), $Si_3N_4$ (n = 2.01 at 600 nm), and GaP (n = 3.37 at 600 nm), which span a wide range of refractive indices. In the second approach, tunability was demonstrated by adjusting the cavity thickness in the design, with the entire cavity composed of $TiO_2$. Again, for the proof-of-concept demonstrations, three different cavity thicknesses of 80, 200, and 270 nm were selected. Additionally, numerical and experimental Fano spectral line shapes were obtained for electric field polarization along and perpendicular to the grating direction. The engineered structure exhibits an outstanding range of tunability while preserving narrow Fano linewidths in all designs, spanning 194 nm across the visible (from 500 to 694 nm). In addition, the pronounced Fano resonances exhibit a maximum Q-factor of 217 in the visible region, which arises from a field localization into the cavity of around 26 folds that of the applied source. We numerically analyzed and demonstrated the electric field distribution in the cavity region, as well as the mode decomposition, to understand the resonant mechanism underlying the excited resonances. Here, the experimental realization of bright and vivid structural colors from all of these designs opens a new direction for various possible photonic applications. Also, importantly, the straightforward etch-free fabrication process used in this work reduces fabrication complexity and process-induced optical degradation.

## RESULTS AND DISCUSSION

The structural design of the proposed Fano resonance-based metasurface is presented in Figure 1a, where a one-dimensional periodic grating layer is integrated with a dielectric multi-layered cavity. The grating material is the low refractive index transparent polymer poly(methyl methacrylate) (PMMA) with a fixed height of 100 nm and an edge-to-edge distance of 250 nm. As a completely transparent material within the visible spectrum, PMMA exhibits good stability against intense light sources, including high-power laser beams[49]. The cavity consists of three layers of dielectric materials, with the middle layer chosen to be 50 nm thick silicon nitride ($Si_3N_4$), sandwiched between two layers of $TiO_2$, each 110 nm thick. The optical constant of cavity materials and the thickness of the cavity control the position of the Fano resonances. Therefore, by changing the material of the cavity's middle layer and overall thickness, we show the controlled tuning of Fano resonance. The materials of the device, $TiO_2$ and GaP, were deposited by RF sputtering, $Si_3N_4$ was deposited by plasma-enhanced PECVD, and PMMA was deposited by spin-coating. The grating patterns were created using electron beam lithography (EBL), and the entire device was fabricated on a fused silica ($SiO_2$) substrate. Importantly, the complete device fabrication procedure does not require any etching (either dry or wet), reducing fabrication process-induced complexity and optical degradation. Figure 1b represents the top view of a scanning electron microscopy (SEM) image from the fabricated device, and the detailed fabrication procedure is discussed in the method section and Supporting Information S1.

The Fano resonance arises due to the coupling between two oscillating states with strongly different damping rates[2], where the phase at resonance changes by $\pi$, resulting in a sharp asymmetric line shape. The interference between the radiative grating mode (bright state) and the subradiative cavity mode (dark state) is well known to yield a sharp Fano resonance[25]. The periodic in-plane momentum of the grating supports bright modes that couple efficiently to free-space radiation, while the cavity supports vertically confined, momentum-mismatched guided modes (continuum state) that are intrinsically subradiant. The Fano resonance line curve can be represented by the following scattering formula[4,16]:

$$P(\omega) = P_0 + R_0 \frac{[q+2(\omega-\omega_0)/\Gamma]^2}{1+[2(\omega-\omega_0)/\Gamma]^2} \tag{1}$$

where $\omega_0$ denotes the resonant frequency, $\Gamma$ represents the resonant dip linewidth, $P_0$ is the background scattering parameter, $R_0$ stands for the coupling coefficient, and q is the ratio of the resonant state to the continuum state, describing the asymmetric parameter of the resonant spectrum. For $q >> 1$, the discrete resonant state dominates the line shape with the conventional symmetric Lorentzian profile. In the case of $q \sim 1$, when the discrete resonant state is comparable with the continuum state, the interference effect is more prominent, resulting in an asymmetric Fano line shape. While $q = 0$, the Fano resonance line shape turns to a symmetric dip, occasionally referred to as an antiresonance.

We numerically tuned the refractive index of the middle layer of the cavity from 1.0 to 5.0 as a function of the wavelength to achieve the Fano resonance in the reflectance spectra. All numerical analyses of the reflectance spectra and electric field map distribution were performed using the commercially available finite-difference time-domain (FDTD) software Lumerical. Figures 1c and 1d show the tuning of Fano resonances as a function of the wavelength and refractive index of the middle layer. Both figures demonstrate the continuous change in the reflectance spectrum distribution of dark and bright regions, which is attributed to the Fabry-Pérot cavity mode effect due to the dielectric cavity layers. There is no Fano resonance line shape in the complete dark region, as the contribution of continuum states from the cavity layers is minimal. Figure 1c depicts three resonant spectral lines that red-shift in accordance with the increments of the rising refractive index. The resonant spectral line in the longer wavelength region displays the Fano line shape in the first bright region from the refractive index of 1.0. However, the intensity decreases while approaching the dark region and vanishes completely when $n > 2.5$. Another resonant spectral line starts from 552 nm when $n = 1.0$ and exhibits a Lorentzian profile ($q >> 1$) while $n = 1.30$ to $n = 2.30$ with a quality factor of 433 while residing in the dark region (Supporting Information S2), however, gradually acquiring a Fano line shape ($q \sim 1$) while approaching the bright region. The third resonant spectral line, which appears in the shorter wavelength region, features a Fano line shape for materials with a higher refractive index value (Supplementary Information S2). Similarly, Figure 1d presents two resonant spectral lines with reflectance gradient and Fano line shape, where the red shift is more rapid compared to Figure 1c. To show the Fano resonance tuning experimentally, we chose three different materials – $Si_3N_4$, PMMA, and GaP as a middle layer of the cavity.

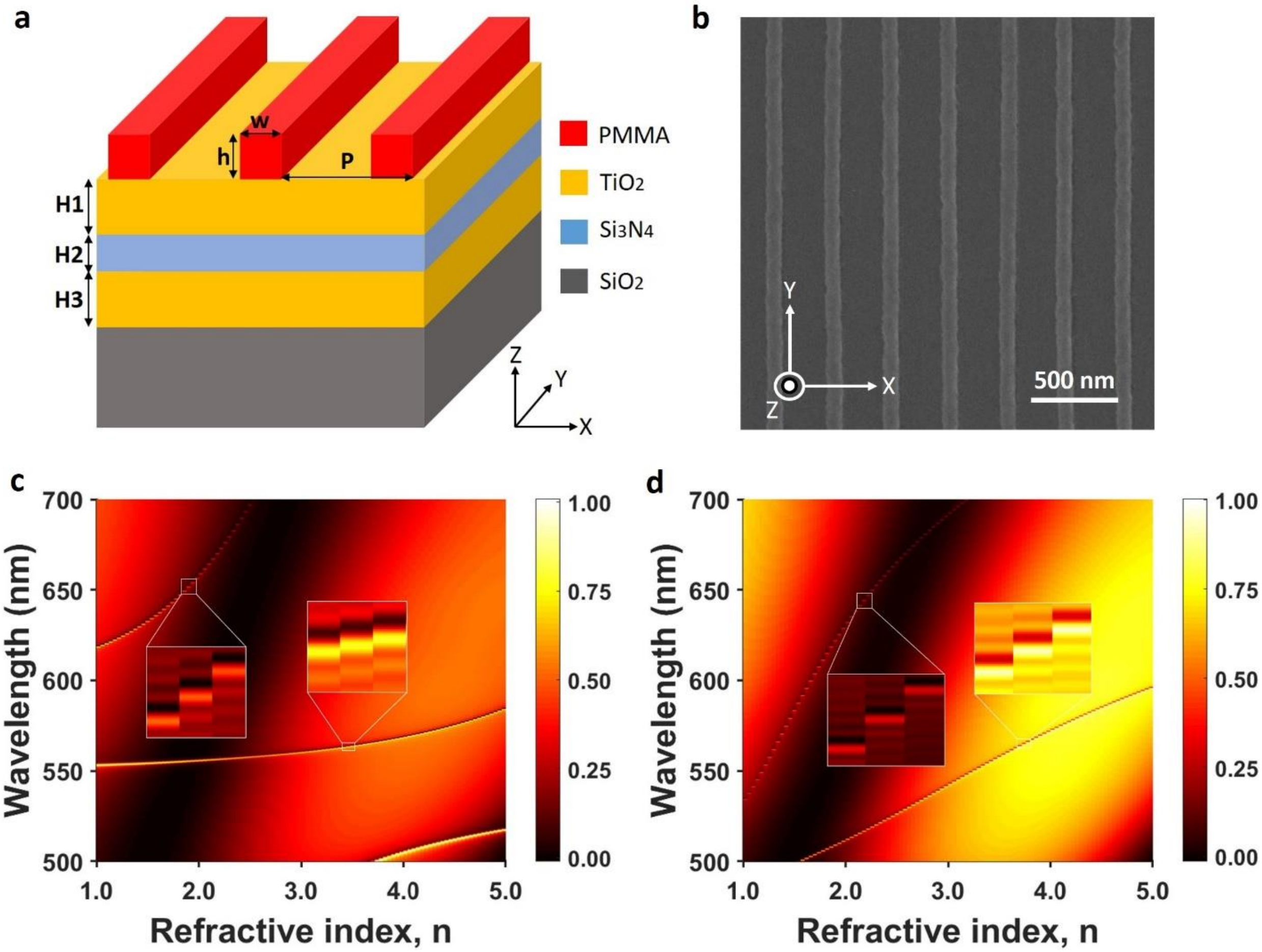

**Figure 1.** Structural design, SEM image, and Fano resonance tuning. (a) Schematic illustration of the metasurface consists of a dielectric grating and a cavity. The grating material is PMMA with a period (P) of 330 nm, a height (h) of 100 nm, and a width (w) of 80 nm. The cavity consists of three layers, where the top (H1) and bottom (H3) layers are $TiO_2$ of thickness 110 nm, and the middle (H2) layer is $Si_3N_4$ with a thickness of 50 nm. (b) Top-view of the scanning electron microscopy (SEM) image of the fabricated metasurface. (c, d) Numerical spectral modulation of reflectance as a function of the real value of refractive index, n, for the middle layer of the cavity from 1.0 to 5.0, where the electric field of the applied source is polarized to the y-axis (c) and x-axis (d). The insets show the magnified images of resonances where the sharp change from bright field to dark field denotes Fano resonances.

The reflectance performance, along with the angle dependence on the applied source and the electric field distribution within the cavity region, is shown in Figure 2. Figures 2a and 2b demonstrate the numerical reflectance spectra for the electric field polarization directions of the applied source along the y-axis and x-axis, where the Fano resonance dips appear at 660.1 and 633.6 nm, respectively, with a Q-factor ($\omega_0/\Gamma$) of 1,000 and 1,006. The experimental reflectance of the corresponding Figures 2a and 2b is presented in Figures 2d and 2e with relatively lower Q-factors of 87 and 105, where the Fano resonance appears at 660 and 622 nm, respectively. A homemade optical setup was built to characterize the experimental reflectance based on the back-focal-plane (BFP) imaging system (see Method section and Supporting Information S3). The relatively lower Q-factor observed in the experimental results is attributed to the small metasurface area (50 $\mu m^2$), imperfections in the fabricated samples, and limitations of the spectrograph in the characterization setup. Figures 2c and 2f demonstrate the experimental results of angle-resolved reflectance, where the sharp changes in reflectance intensity denote the Fano resonances. The resonances change the spectral positions very slightly while maintaining the Fano line shape within a wide angular range, indicating the stability of the Fano resonance to the applied source.

The numerical electric field distribution calculated on the vertical cross-section of metasurfaces is depicted in Figures 2g and 2h at the Fano resonance dip points of the corresponding Figures 2a and 2b. The electric field is entirely accumulated within the cavity region (Figure 2g) and densely concentrated within the middle layer of the cavity (Figure 2h), where the electric field is boosted by 14 folds. In comparison, the electric field accumulation within or around the grating is almost negligible in both Figures 2g and 2h. This enhancement of the electric field within the cavity is crucial, for example, increasing the optical gain and lasing from quantum emitters (such as quantum dots and wells) that can be embedded within the cavity. We also investigated the responsible modes for the Fano resonances from the electric field distributions of the horizontal and vertical cross-sections of the cavity (Supporting Information S4), which are electric and magnetic dipoles, respectively, with the applied source polarization directions along the y- and x-axis, respectively. The numerical and experimental reflectance spectra, the experimental angle-resolved reflectance, and the cross-sectional electric field distribution for the middle layers of the cavity with PMMA and GaP are presented in the Supporting Information (S5 and S6) to further demonstrate Fano resonance tuning across different materials.

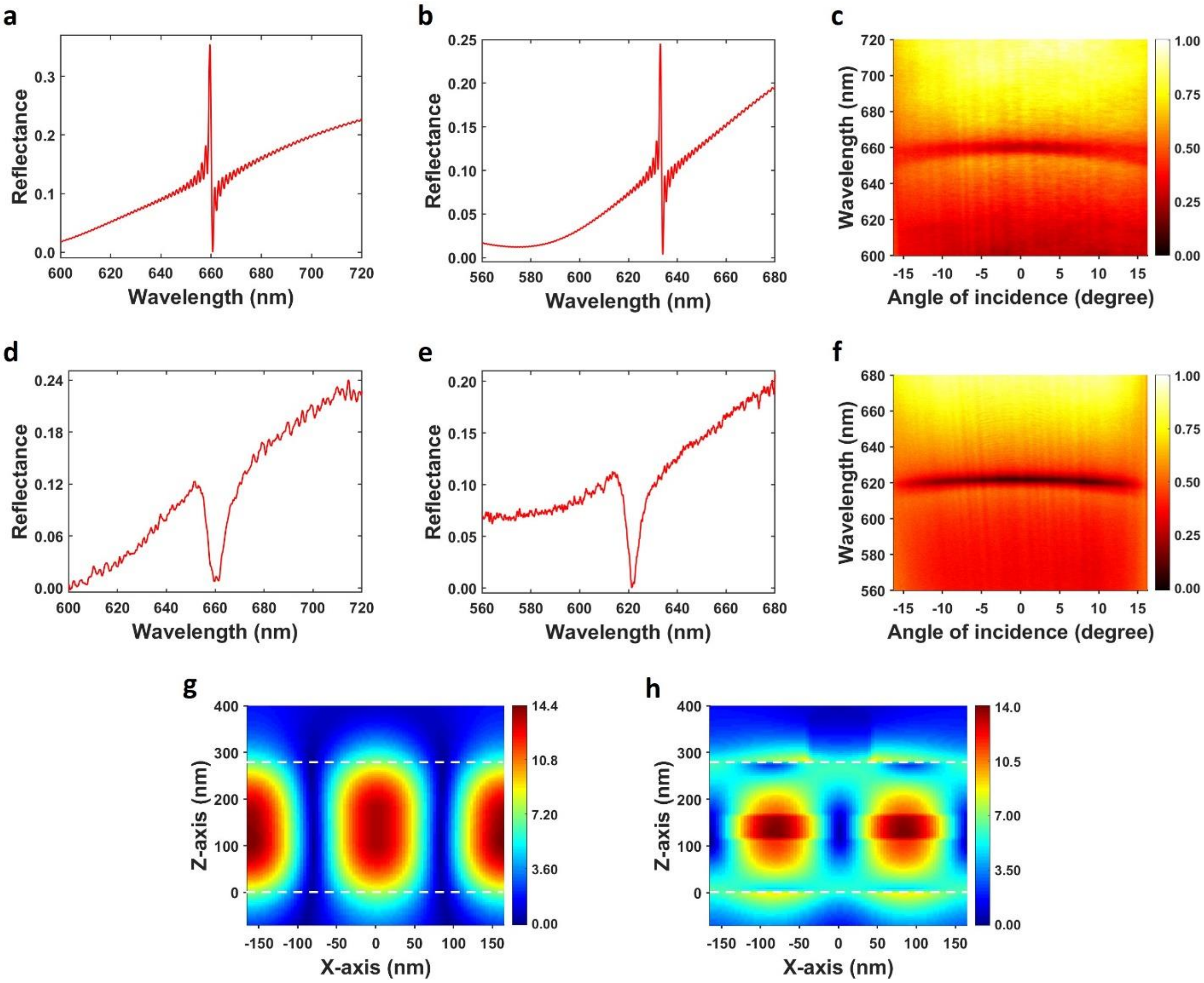

**Figure 2.** Reflectance spectra, angle-resolved reflectance, and electric field maps from the vertical cross-section where the middle cavity layer is $Si_3N_4$. (a, b, d, e) Numerical (a, b) and experimental (d, e) reflectance spectra for the electric field polarization of the applied source along the y-axis (a, d) and along the x-axis (b, e), where the Fano dips at 660.1 (a), 633.6 (b), 660.0 (d), and 622.0 nm (d), respectively. (c, f) Normalized experimental angle-resolved reflectance for the electric field polarization of the source along the y-axis (c) and x-axis (f), denoting the wide-angle stability of the Fano resonance. (g, h) Numerical electric field map from the vertical cross-section of the metasurface at the resonance dip points of 660.1 (a) and 633.6 nm (b), respectively, denoting the light energy completely trapped into the cavity region (g) and heavily concentrated into the middle layer of the cavity region (h) [Here dashed lines denote the cavity region].

Besides tuning the Fano resonance by changing the materials of the middle layer of the cavity, it can be further tuned by modulating the thickness of the cavity materials. To prove this numerically, we consider the entire cavity as a $TiO_2$ layer and vary its thickness from 20 to 270 nm. The results are shown in Figures 3a and 3d, where the resonant spectral position undergoes a rapid red shift as the cavity thickness increases. However, the resonant spectra show Fano line curves only in the bright region of Figures 3a and 3d, where $q \sim 1$. The insets from the bright regions exhibit the rapid changes in reflectance, denoting the Fano resonances. To demonstrate this experimentally, we selected three different cavity thicknesses: 80, 200, and 270 nm. The resonant spectra in the dark region exhibit Lorentzian profiles ($q >> 1$). Figures 3b and 3e represent the numerical reflectance curves of Lorentzian profiles for the thicknesses of cavities 140 and 250 nm from Figure 3a, where the spectral position and Q-factor of resonances are 618 nm and 520 (Figure 3b), and 544 nm and 550 (Figure 3e), respectively. Likewise, Figures 3c and 3f also present the numerical Lorentzian reflectance profiles for the cavity thicknesses of 120 and 230 nm, as shown in Figure 3d, with a spectral position at 537 nm and a Q-factor of 610 (Figure 3c) and 494 nm and a Q-factor of 716 (Figure 3f).

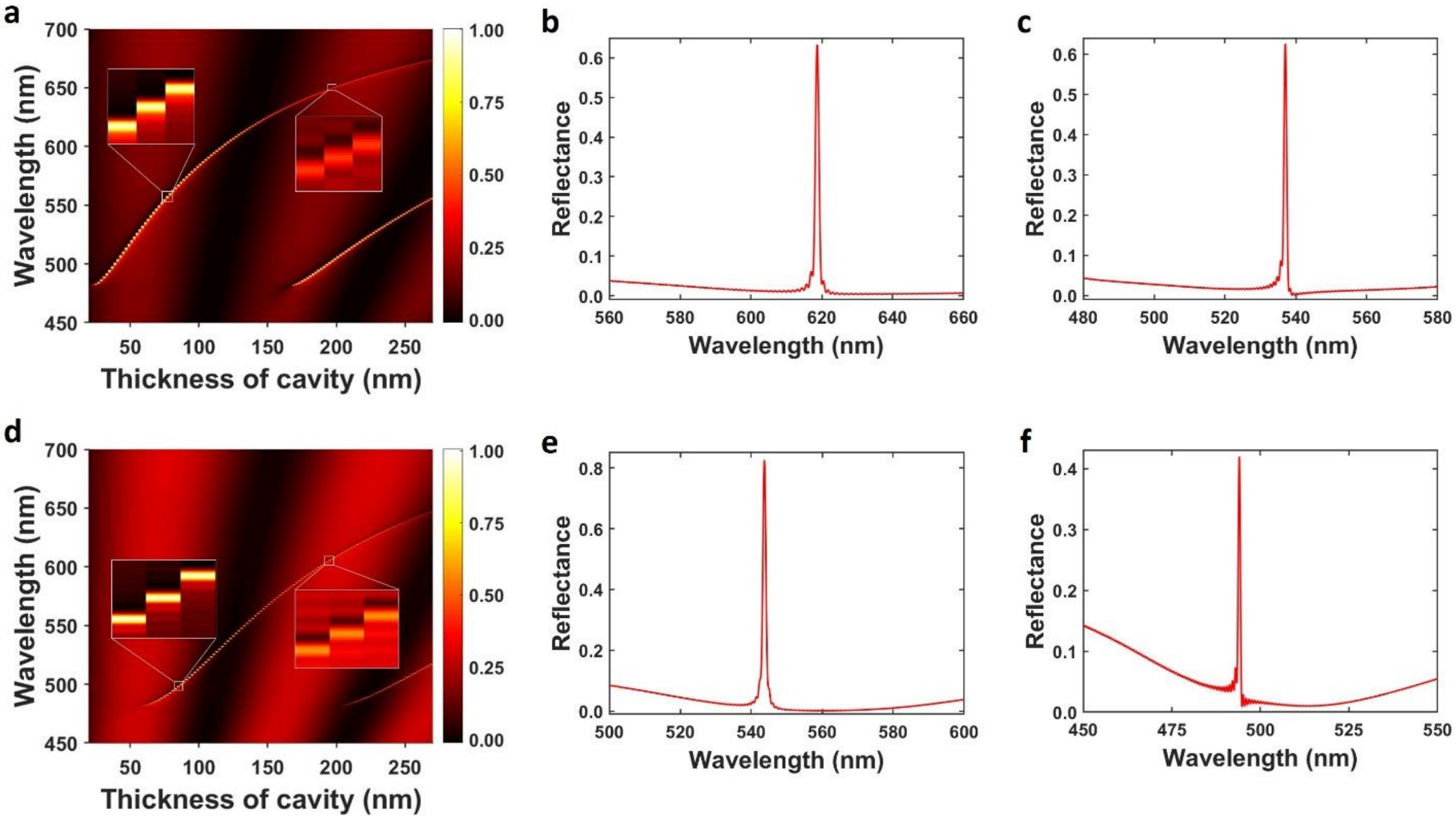

**Figure 3.** Tuning of resonances by changing the design. (a, d) Numerical reflectance as a function of the cavity thickness for the electric field polarization of the applied source along the y-axis (a) and the x-axis (d). The insets show a magnified view of the resonances, where the sharp change from bright to dark field denotes the Fano resonance.

(b, e) Numerical reflectance spectra for the cavity thickness of 140 (b) and 250 nm (e) from the dark region of (a), exhibiting Lorentzian resonances. (c, f) Numerical reflectance spectra for the cavity thickness of 120 (c) and 230 nm (f) from the dark region of (d), also featuring Lorentzian resonances.

The numerical and experimental reflectance, the experimental angle-resolved reflectance, and the numerical electric field distribution map from the vertical cross-section of cavities are presented in Figure 4 for a cavity thickness of 80 nm. Figures 4a and 4d display the numerical and experimental reflectance results for the electric field polarization direction of the applied source along the y-axis. Both Figures represent the ideal Fano resonance line shape, with the Fano spectral positions and Q-factors at 563 nm and 286 (Figure 4a), and 570 nm and 120 (Figure 4d), respectively. Similarly, Figures 4b and 4e demonstrate the numerical and experimental reflectance when the polarization direction is along the x-axis, with Fano spectral positions and Q-factors of 496 nm and 533 (Figure 4b) and 500 nm and 213 (Figure 4e), respectively. The slight difference between the numerical and experimental Fano spectral positions is attributed to fabrication imperfections relative to the ideal numerical structure. The experimental angle-resolved reflectance (Figures 4c and 4f) demonstrates that the spectral positions of the Fano resonances change very little over a wide range of angles for the applied light source. However, the Fano resonance is more angle invariant, pronounced, and intense in Figure 4c compared to Figure 4f. The numerical electric field map analysis reveals that the light energy is heavily localized (~18 folds) in the cavity region (Figure 4g), with the electric field polarization angle aligned along the y-axis. However, the light energy is concentrated heavily (~26 folds) outside of the cavity region (Figure 4h), while the electric field polarization angle is along the x-axis. The numerical and experimental reflectance spectra, the experimental angle-resolved reflectance, and the cross-sectional electric field distribution, with the thickness of the $TiO_2$ cavity layers set at 200 and 270 nm, are presented in the Supporting Information S7 and S8, respectively, to provide further insight into the thickness tuning.

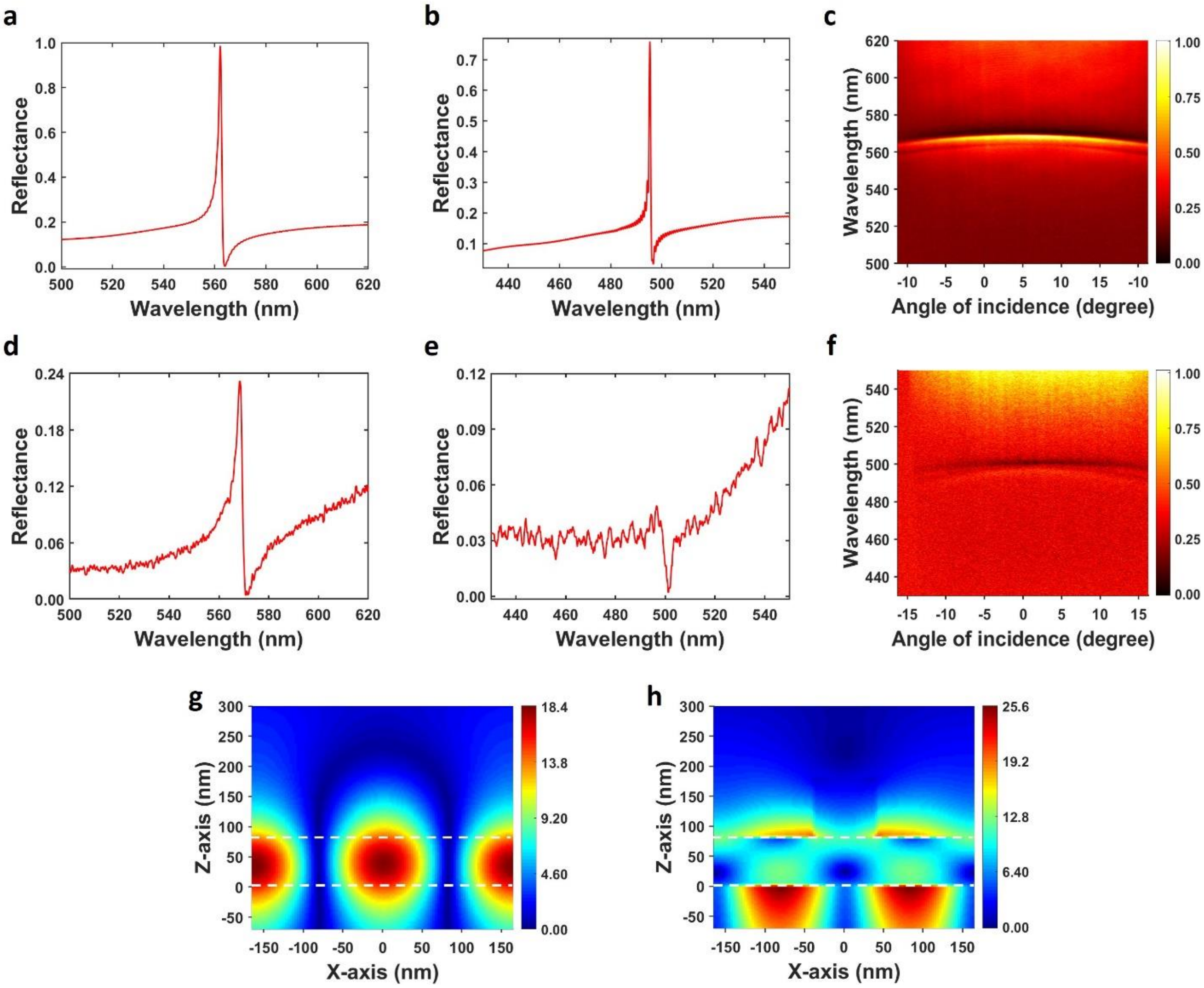

**Figure 4.** Reflectance spectra, angle-resolved reflectance, and electric field maps from the vertical cross-section where the cavity layer is $TiO_2$ with a thickness of 80 nm. (a, b, d, e) Numerical (a, b) and experimental (d, e) reflectance spectra for the electric field polarization of the applied source along the y-axis (a, d) and along the x-axis (b, e), where the pronounced Fano dips at 562.6 (a), 495.6 (b), 570.0 (d), and 501.0 nm (d), respectively. (c, f) Normalized experimental angle-resolved reflectance for the electric field polarization of the source along the y-axis (c) and x-axis (f), demonstrating the wide-angle stability of Fano resonance similar to the multi-dielectric cavity. (g, h) Electric field map from the vertical cross-section of the metasurface at the resonance dip point of 562.6 (a) and 495.6 nm (b), respectively, denoting the light energy completely trapped into the cavity region (g) and outside of the cavity region (h) [Here dashed lines denote the cavity region].

As the electric field trapping and mode generation are completely dictated by the cavity layers, the resonant spectral positions are almost invariant with respect to the geometry of the grating. However, the Fano line shape changes based on the width of the grating. To demonstrate this numerically, we varied the grating width from 20 to 270 nm while retaining the thickness fixed at 100 nm. Figures 5a and 5b demonstrate the modulation of grating width, where the resonant spectral positions are invariant in Figure 5a and vary slightly in Figure 5b. The insets (Figures 5a and 5b) show that the ideal Fano line shapes appear between 40 and 120 nm, where the contrast between bright and dark is more pronounced. We chose a grating width of 80 nm (Figure 1a) for our device design, as the Fano line shape appears perfectly, and the field accumulation in the cavity region is maximum at this width. After the grating width reaches 120 nm, the Fano line shapes gradually transition to Lorentzian line shapes, as discussed in Supporting Information S9. The experimental realization of grating width modulations is presented in Figures 5c-5h, featuring minimal variations in structural colors, which were photographed with a smartphone camera and an optical microscope. Different shades of red were photographed (Figure 5c) across the various grating widths, indicating the narrow spectral-range tuning of the Fano resonances, while the middle layer is $Si_3N_4$. Similarly, different shades of orange and yellow were photographed, while the middle layers of materials were PMMA (Figure 5d) and GaP (Figure 5e). Figures 5f to 5h show photographs of different shapes of structural colors in orange, green, and magenta, corresponding to cavity thicknesses of 80 nm (Figure 5f), 200 nm (Figure 5g), and 270 nm (Figure 5h), respectively. If we compare the standard colors and their associated wavelengths, we find that all colors correspond to the Fano resonance wavelengths.

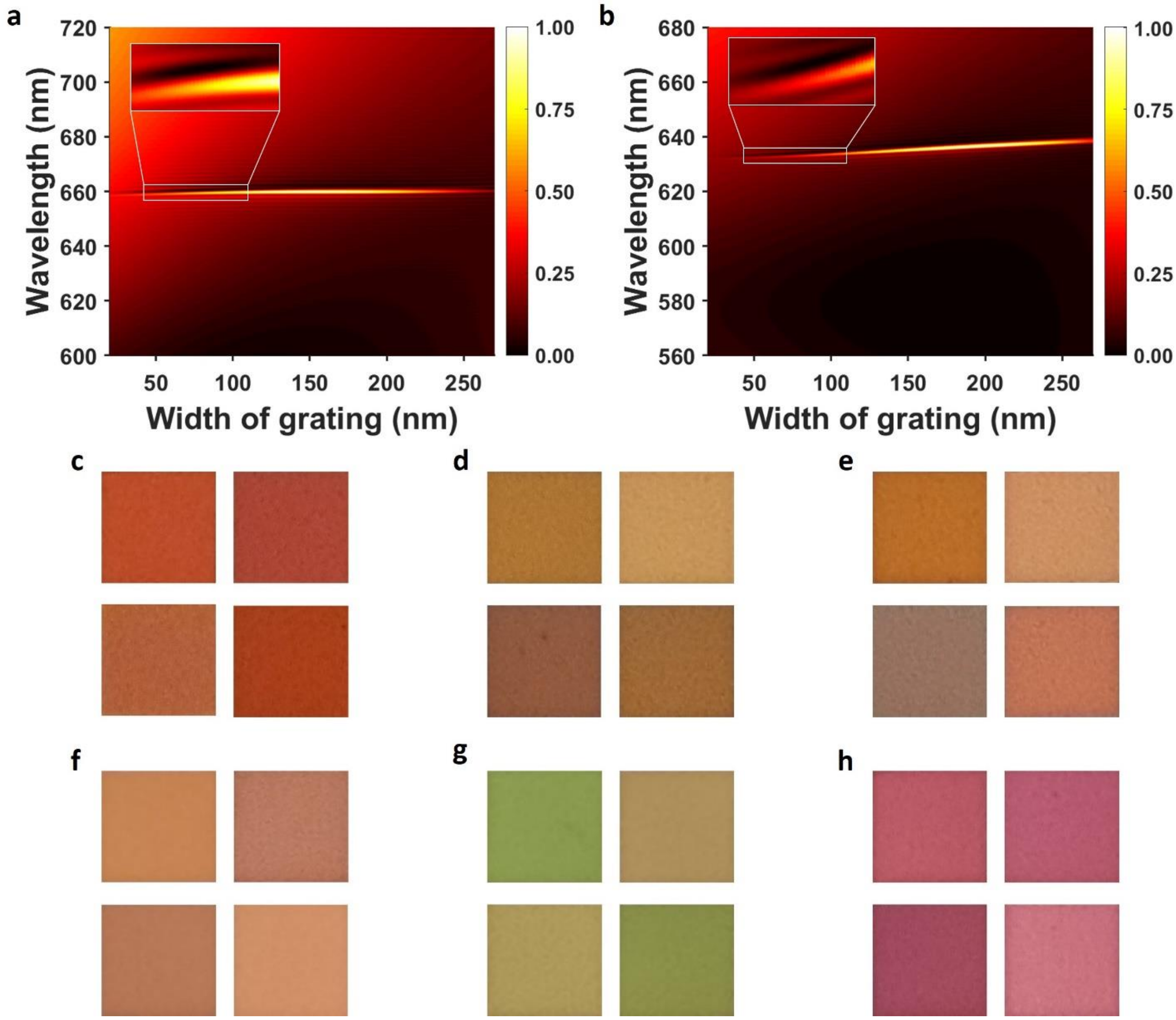

**Figure 5.** Tuning of grating width and structural colors. (a, b) Numerical reflectance as a function of the width of the grating for the electric field polarization of the applied source along the y-axis (a), and the x-axis, demonstrating a pronounced asymmetric Fano linewidth between 40 and 120 nm (insets). (c – e) Experimental photographs of structural colors for the unpolarized light source and through the 10× objective of an optical microscope from the square metasurface of side 50 μm, while the material of the middle layers is $Si_3N_4$ (c), PMMA (d), and GaP (e). (f - h) Experimental photographs of structural colors from the metasurface of cavity material $TiO_2$, while the thickness of the cavity is 80 (f), 200 (g), and 270 nm (h), respectively. The marginal variation in color is due to variation in the grating width.

## CONCLUSIONS

In summary, we have proposed and demonstrated both numerically and experimentally a multi-layered dielectric metasurface architecture specifically for the visible range, where a cavity layer supports vertically confined waveguide modes that are intrinsically sub-radiant, and the grating layer supports radiating discrete resonant states. The controlled interference and hybridization between these two modes effectively lead to the realization of sharp Fano resonances. By strategically adjusting the cavity parameters, the proposed device enables wide tunability of the Fano resonance spectrum, spanning 194 nm within the visible spectrum. Besides, the high concentration of light in the cavity region enables new opportunities for low-loss nonlinear photonic devices in the visible. Additionally, the proposed device architecture maintains the Fano linewidth across the visible spectrum even with the use of a very thin $TiO_2$ cavity layer, enabling highly compact photonic devices. Furthermore, the experimental presentation of characteristic structural colors of Fano resonances in the visible paves the way possibly for metasurface-based digital display and printing technologies. Finally, this etch-free planar device could be a compelling platform for Fano-resonance-based next-generation optics and photonics technologies in the visible.

## METHODS

**Numerical calculations.** The numerical reflectance calculations and all cross-sectional electric field map images were performed using the finite-difference time-domain (FDTD) based commercial software, Lumerical. We applied a fine mesh of 4 nm to the device. We also employed the periodic and perfectly matched layer (PML) boundary conditions along the horizontal and vertical directions to replicate the periodic structure and absorbing layers. A normally applied source of plane waves with proper polarization direction, which is Bloch/periodic type, was chosen as the light source. The refractive indices of $SiO_2$ and $Si_3N_4$ were chosen from the material database of the software; the refractive index of PMMA was chosen to be 1.49, while for $TiO_2$ and GaP, ellipsometer-calculated optical constants were imported into the software.

**Fabrication.** Initially, the standard piranha-cleaned and dry-fused silica (SiO2) wafers were placed in the RF sputter chamber under high vacuum conditions to deposit a 110 nm thick $TiO_2$ film. Then, a plasma-enhanced chemical vapor deposition (PECVD), a spinner, and an RF sputter were used to deposit $Si_3N_4$, PMMA, and GaP, respectively, for the middle layer. Then another 110 nm thick layer of $TiO_2$ was deposited using RF sputtering. After depositing the PMMA (e-beam resist) with a thickness of 100 nm on top of multi-dielectric cavity layers, a very thin layer (7 nm) of electrically conductive polymer Espacer 300Z (SHOWA DENKO) was spin-cast on top of the PMMA to avoid charge accumulation during the electron beam lithography (EBL) process. After that, the EBL (FEI Nova NanoSEM 600) was conducted on the PMMA, followed by the cleaning of the Espacer with DI water (1 min), then the development with MIBK: IPA (1:3) (1 min), IPA (20 s), and DI water (20 s). The remaining PMMA pattern after the development was the target grating structure.

**Optical characterization.** A back-focal plane (BFP) imaging setup integrated with a slit spectrometer and a CCD camera was employed to measure the reflectance spectra. The BFP setup includes a 4f lens assembly with several mirrors, a halogen lamp for illumination, a Mitutoyo Plan Apo 10× objective (NA 0.28), a polarizer, and a spatial filter. The structural colors were photographed using a smartphone camera (Samsung Galaxy A31) through a bright-field illuminated ZESIS (Axio Scope AI, 10X objective, NA 0.2) microscope.

## AUTHOR INFORMATION

### Corresponding Authors

**Hilmi Volkan Demir** – LUMINOUS! Centre of Excellence for Semiconductor Lighting and Displays, School of Electrical and Electronic Engineering, The Photonics Institute (TPI), Nanyang Technological University, 639798, Singapore; School of Physical and Mathematical Sciences, Division of Physics and Applied Physics, Nanyang Technological University, 639798, Singapore; Department of Electrical and Electronics Engineering and Department of Physics, UNAM - Institute of Materials Science and Nanotechnology, Bilkent University, Ankara 06800, Turkey; orcid.org/0000-0003- 1793-112X; Email: volkan@bilkent.edu.tr, hvdemir@ntu.edu.tr

### Authors

**Md Rumon Miah -** Department of Electrical and Electronics Engineering and Department of Physics, UNAM - Institute of Materials Science and Nanotechnology, Bilkent University, Ankara 06800, Turkey; orcid.org/0000-0001-5247-3967

### Notes

The authors declare no competing financial interest.

## ACKNOWLEDGMENTS

The authors gratefully acknowledge the financial support from TUBITAK 20AG001 and 121C266. H. V. D. also acknowledges the support from TUBA.

## REFEFENCES

# Supporting Information for

## Table of contents

## S1. Device fabrication workflow

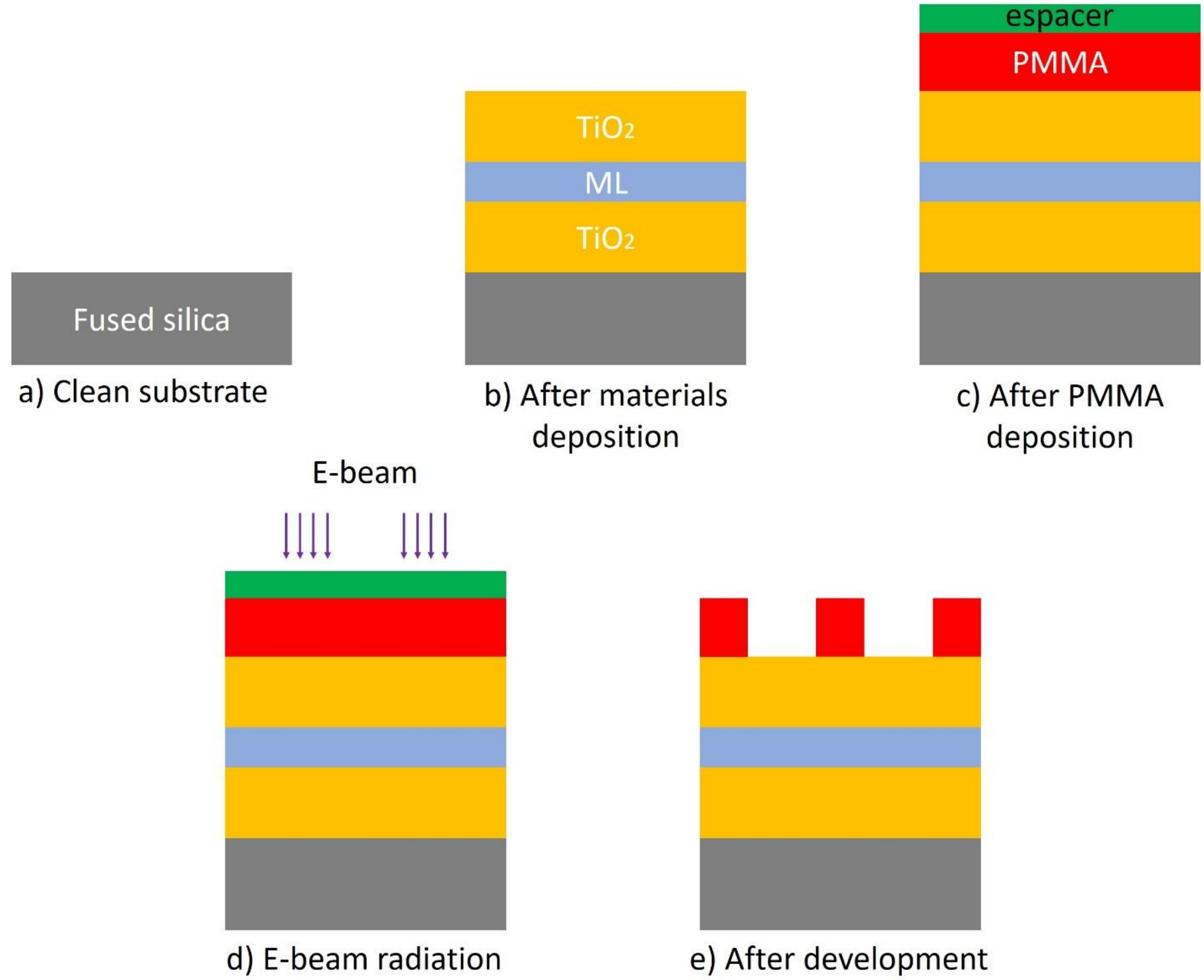

**Figure S1.** Schematic of device fabrication workflow. (a) Piranha ($H_2SO_4$ and $H_2O_2$) cleaned fused silica substrate. (b) Layer-by-layer materials deposition, where MI means Middle Layer, which can be $Si_3N_4$, PMMA, or GaP. (c) Spin-casting of PMMA and espacer for the EBL. (d) Radiation of E-beam. (e) Final structure obtained after washing the espacer and developing the PMMA.

## S2. Numerical reflectance of Lorentzian and Fano shapes

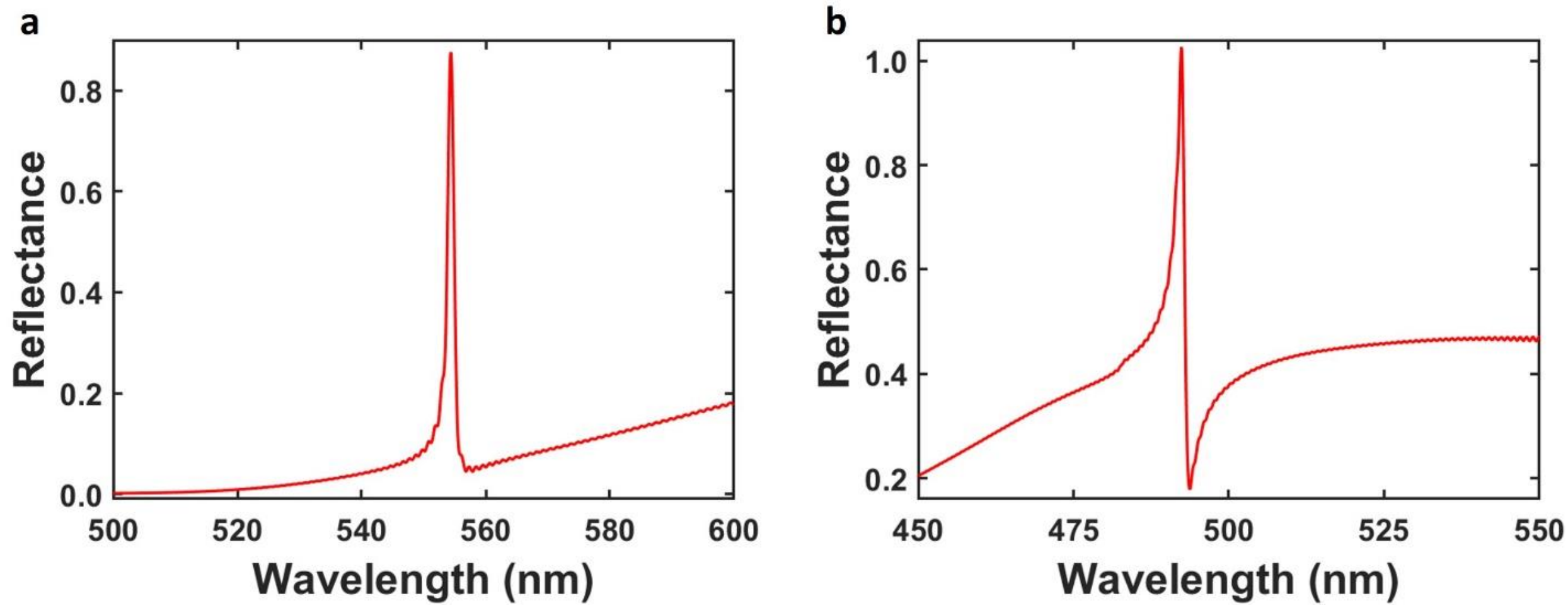

**Figure S2.** Numerical reflectance spectra from Figure 1c. (a) Reflectance from the first dark region of Figure 1c, with the refractive index value of the middle layer is 1.49 (PMMA). (b) Reflectance for the third resonant spectral curve (lower wavelength region) from Figure 1c, with the material of the middle layer being silicon (Si).

To further explain Figure 1c, as discussed in the main text, we present two numerical reflectance calculations. The second resonant spectral line in Figure 1c shows the Lorentzian line shape ($q >> 1$) for $n = 1.30$ to $n = 2.30$ (first dark region). Figure S2(a) demonstrates the Lorentzian profile of reflectance at 554.2 nm with the full-width at half-maximum (FWHM) value of 1.28 nm and Q-factor of 433. On the other hand, Figure S2(b) represents the Fano-shaped reflectance from the third resonant spectral line from Figure 1c, with a high refractive index material of silicon (Si) being the middle layer.

## S3. Back-focal-plane (BFP) imaging setup

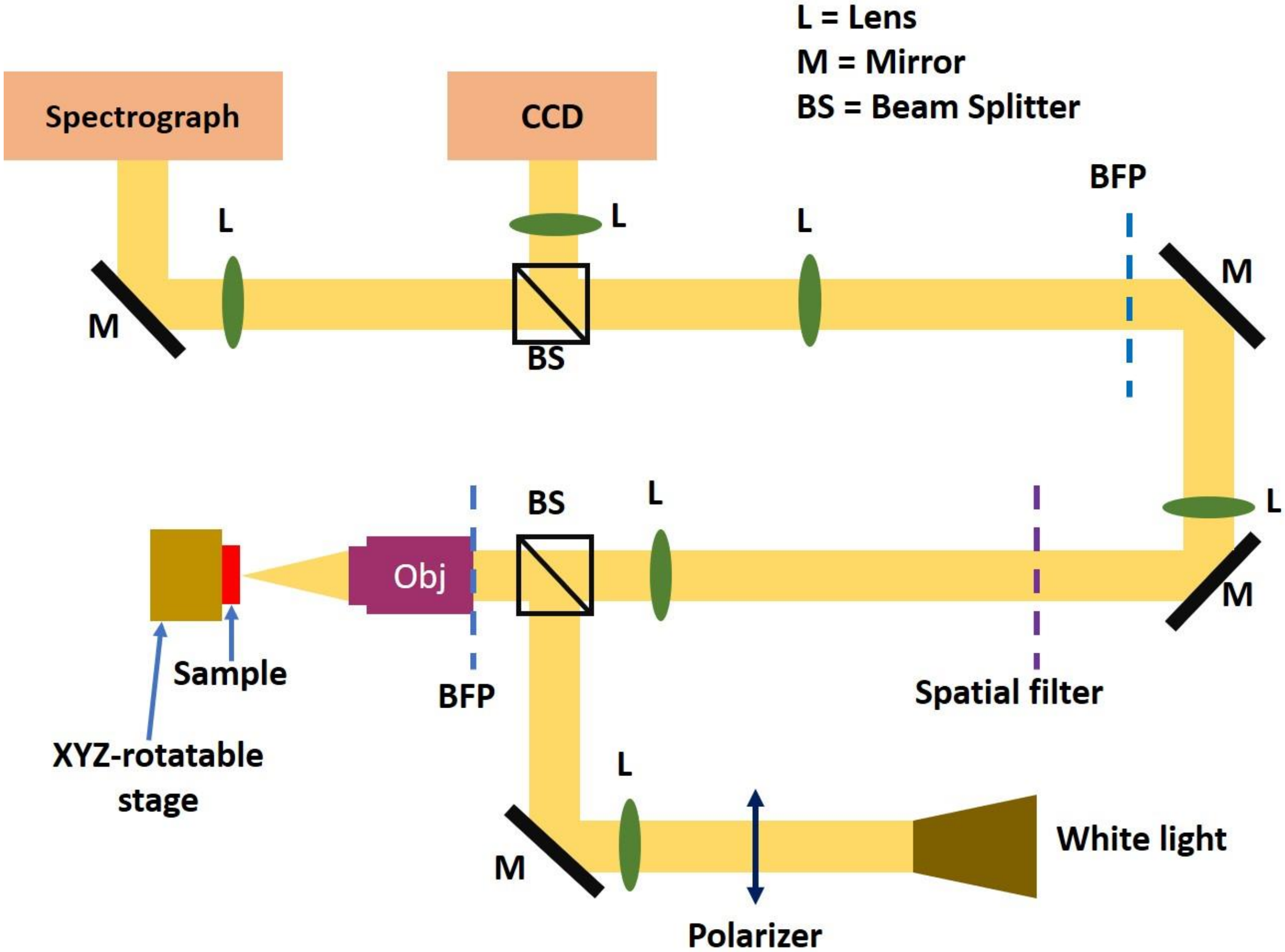

**Figure S3.** Schematic of a home-made BFP characterization setup, where CCD is the charge-coupled device, and Obj is the objective lens.

## S4. Field distribution and mode decomposition

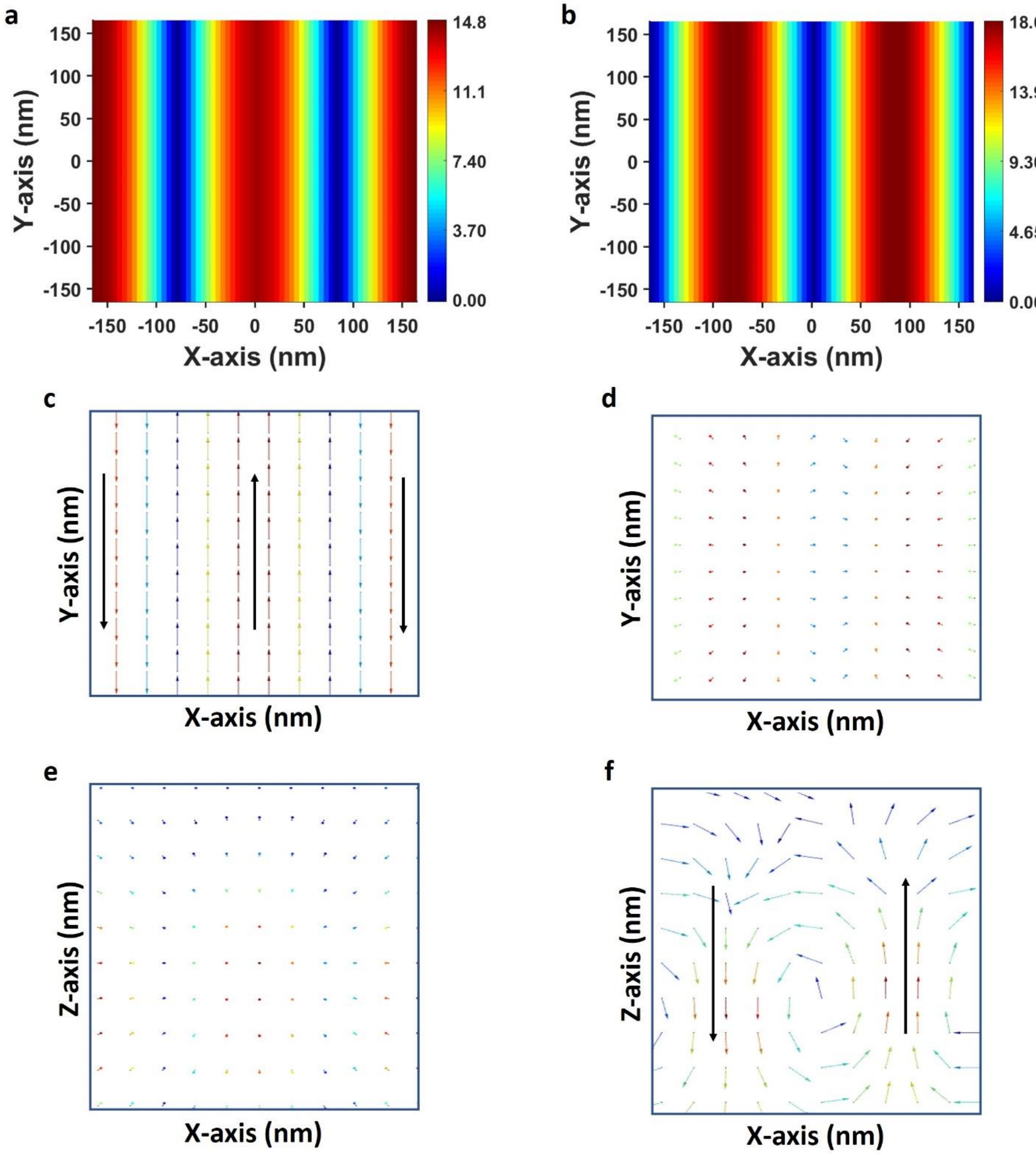

**Figure S4.** Electric field maps and electric field vector distributions. (a, b, c, d) Numerical electric field maps (a, b), and electric field vector distributions (c, d) from the halfway horizontal cross-section of the $Si_3N_4$ middle layer, where the electric field polarization direction of the applied source is along the y-axis (a, c) and along the x-axis (b, d). (e, f) Electric field vector distributions from the vertical cross-section of the device with the $Si_3N_4$ middle layer, where the polarization direction of the applied source is along the y-axis (e) and along the x-axis (f).

To understand the responsible resonant mechanism for the generation of Fano resonances, we investigated the horizontal and vertical cross-sectional electric field map and electric field vector map distributions at the center of the device, where the mid-layer is $Si_3N_4$. By analyzing the field distributions in Figures S4(a), S4(c), and S4(e), which correspond to the Fano resonance point at 660.1 nm from Figure 2a, we can conclude that the excited field within the device is an electric dipole (ED) mode. The linearly distributed electric fields in the device are excited by the linearly polarized electric field of the applied source, which is aligned in the same direction. Similarly, by exploring the field distributions of Figures S4(b), S4(d), and S4(f), which denote the Fano resonance point of 633.6 nm from Figure 2b, we find that the fields into the device are a magnetic dipole (MD) mode, because the excited curvy electric field in the device is coupled with the linearly polarized magnetic field of the applied source. We employed the multipole decomposition analysis technique to further identify the responsible resonances for the Fano line shape within the device. Our analysis is based on the scattering current density induced by the incident light source. The scattering current density is given by

$$\boldsymbol{J} = -i\omega\varepsilon_0(\varepsilon - \varepsilon_d)\boldsymbol{E} \quad \text{(S1)}$$

where $\omega$ is the angular frequency of the wave, $\varepsilon_0$ is the permittivity of vacuum, $\varepsilon$ is the relative permittivity of the device material, $\varepsilon_d$ is the relative permittivity of the surrounding medium, and $\boldsymbol{E} = \boldsymbol{E}(\boldsymbol{r})$ is the electric field. It is possible to identify the excited resonance type based on the Cartesian or spherical coordinates by decomposing the scattering current density[1]. The electric and magnetic dipole moments can be written as

$$\text{Electric dipole (ED) moment: } \boldsymbol{p} = \frac{1}{i\omega}\int \boldsymbol{J} d^3r \quad \text{(S2)}$$

$$\text{Magnetic dipole (MD) moment: } \boldsymbol{m} = \frac{1}{2c} \int [\boldsymbol{r} \times \boldsymbol{j}] d^3 r \tag{S3}$$

where c is the speed of light in vacuum[2,3].

The radiated scattered power from the ED and MD can be expressed as

$$\boldsymbol{P_p} = \frac{k^4 c}{12\pi\varepsilon_0} |\boldsymbol{p}|^2 \tag{S4}$$

$$\boldsymbol{P_m} = \frac{k^4}{12\pi\varepsilon_0 c} |\boldsymbol{m}|^2 \tag{S5}$$

where $k = 2\pi/\lambda$ is the wave number and λ is the wavelength in free space.

## S5. Reflection and electric field distributions for the PMMA middle layer of the cavity

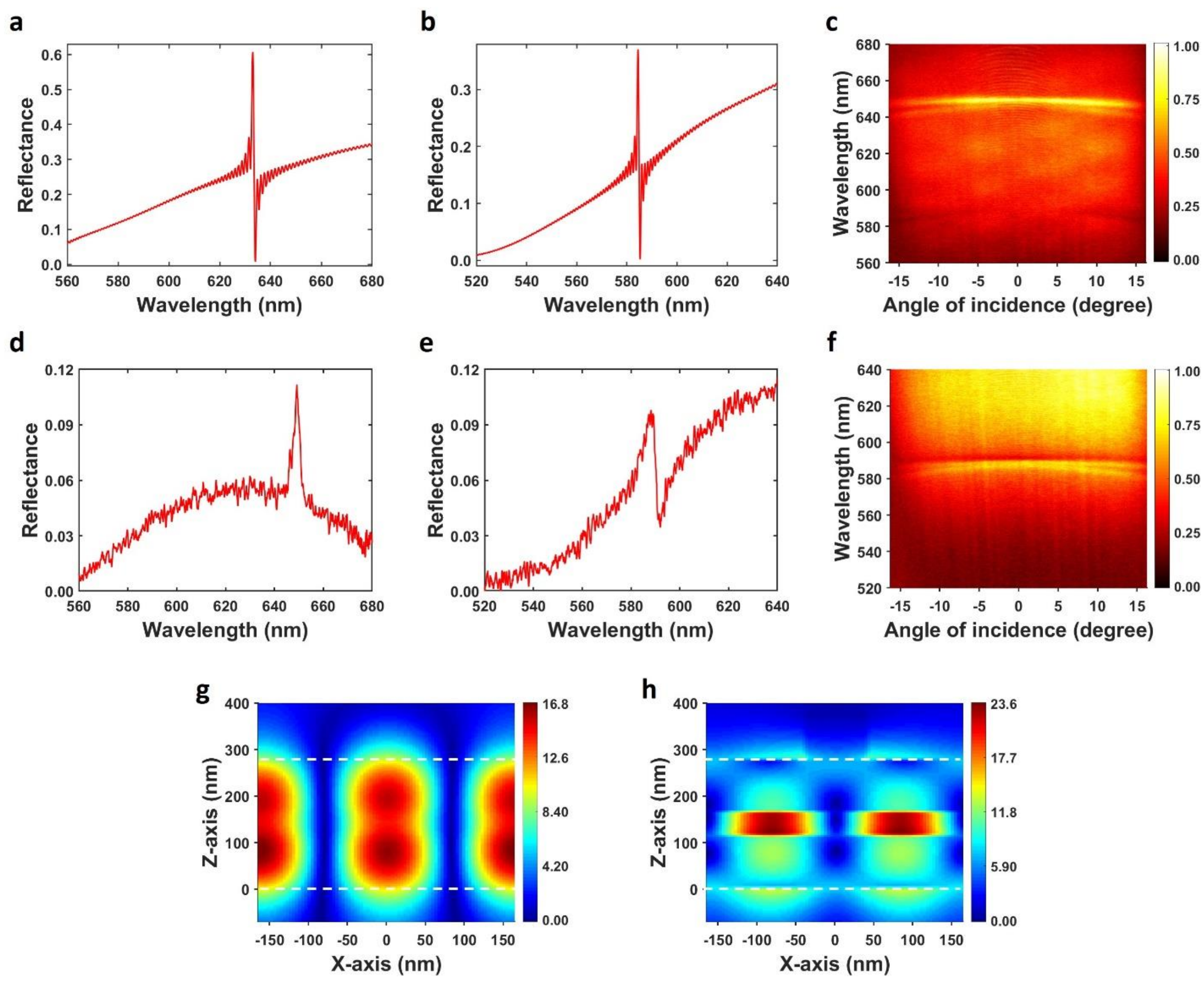

**Figure S5.** Reflectance spectra, angle-resolved reflectance, and electric field maps from the vertical cross-section where the middle layer of the cavity is PMMA. (a, b, d, e) Numerical (a, b) and experimental (d, e) reflectance spectra for the electric field polarization of the applied source along the y-axis (a, d) and along the x-axis (b, e), where the resonance dip points are 633.6 (a), 685.0 (b), and 590.0 nm (e), respectively. However, a Lorentzian shape curve is observed instead of a Fano dip at 650.0 nm (d). (c, f) Normalized experimental angle-resolved reflectance for the electric field polarization of the source along the y-axis (c) and x-axis (f), denoting the wide-angle stability of Fano resonance similar to other cavity materials. (g, h) Electric field maps from the vertical cross-section of the metasurface at the resonance dip points of 633.6 (a) and 685.0 nm (b), respectively, demonstrating the light energy trapped into the $TiO_2$ layers of the cavity region (g) and heavily concentrated into the middle layer of the cavity region (h) [Here dashed lines denote the cavity region].

The reflectance performance, along with angle dependency on the applied source and vertical cross-sectional electric field maps, is demonstrated in Figure S5 for the middle layer of the cavity is PMMA. In Figures S5(a) and S5(b), the Fano resonance dips appear at 633.6 and 585.0 nm, with the FWHMs of 0.64 and 0.62 nm, and Q-factors of 990 and 944, respectively. Figures S5(d) and S5(e) present the experimental reflectance at 650.0 and 590.0 nm, where Figure S5(d) shows the Lorentzian shape resonance, unlike the simulated Fano shape resonance of Figure S5(a). On the other hand, Figure S5(e) demonstrates a Fano line shape similar to its simulated counterpart in Figure S5(b), with a Q-factor of 148. The experimental angle-resolved reflectance is presented in Figures S5(c) and S5(f), where the resonance spectral positions change very slightly. Additionally, the vertical cross-sectional electric field map distributions indicate that the light energy is accumulated more strongly in the cavity region than in the $Si_3N_4$ middle layer cavity. The accumulation is ~17 folds in Figure S5(g), further increasing to ~24 folds in Figure S5(h). This enhanced energy accumulation is due to a large refractive index difference between the PMMA and $TiO_2$ layers.

## S6. Reflection and electric field distributions for the GaP middle layer of the cavity

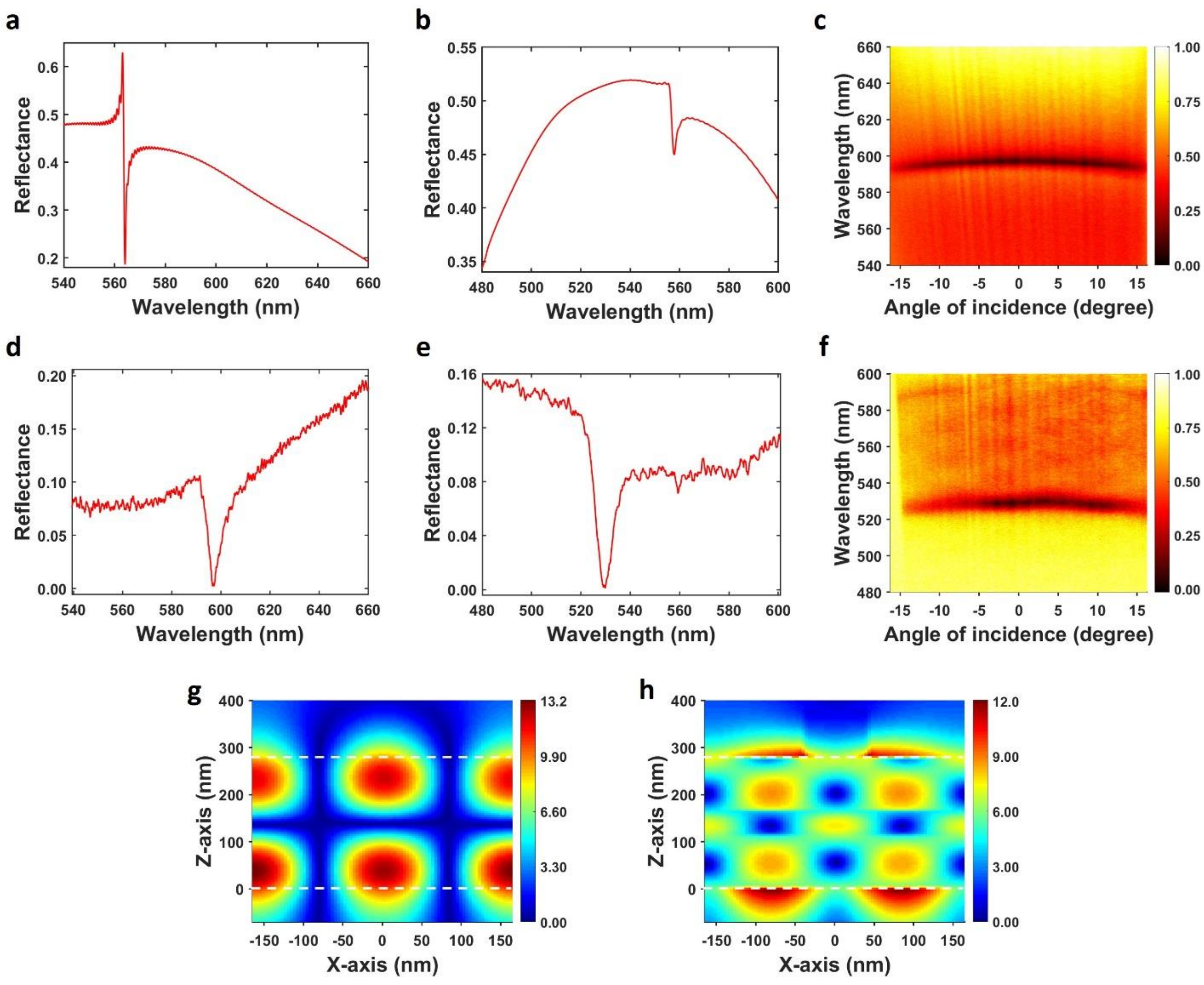

**Figure S6.** Reflectance spectra, angle-resolved reflectance, and electric field maps from the vertical cross-section where the middle layer of the cavity is GaP. (a, b, d, e) Numerical (a, b) and experimental (d, e) reflectance spectra for the electric field polarization of the applied source along the y-axis (a, d) and along the x-axis (b, e), where the resonance dip points are 564.1 (a), 557.7 (b), 596.0 (d), and 530.0 nm (e), respectively, with a significant Fano dip spectral position difference between the numerical and experimental results. (c, f) Normalized experimental angle-resolved reflectance for the electric field polarization of the source along the y-axis (c) and x-axis (f), denoting the wide-angle stability of Fano resonance similar to other cavity materials. (g, h) Electric field maps from the vertical cross-section of the metasurface at the resonance dip points of 564.1 (a) and 557.7 nm (b), respectively, demonstrating the light energy completely trapped into the $TiO_2$ layers of the cavity region (g) and dispersed within and outside of the cavity (h) [Here dashed lines denote the cavity region].

We also investigate the numerical and experimental reflectance with the experimental angle-resolved reflectance, along with the vertical cross-sectional electric field distribution map for a high refractive index material of GaP as the middle layer of the cavity. Figures S6(a) and S6(b) show the numerical reflectance where the Fano resonance dips at 564.1 and 557.7 nm, respectively, with the FWHMs of 0.75 and 1.8 nm, and Q-factors of 752 and 310. Whereas the experimental reflectance is depicted in Figures S6(d) and S6(e) at the spectral positions of 596.0 and 530.0 nm, respectively, with the Q-factors of 96 and 75, respectively. There is a substantial difference in spectral positions and Q-factors between the numerical and experimental results due to the fabrication imperfections and optical constant mismatch. Figures S6(c) and S6(f) show the experimental angle-resolved reflectance, characterized by strong angular stability and a sharp reflection intensity contrast. Additionally, the numerical electric field distribution maps of Figures S6(g) and S6(h) from the vertical cross-section show that the light energy is dispersed into the $TiO_2$ layers and outside of the cavity layers. These electric field distributions differ from the previous two designs, attributed to the high refractive index of the GaP layer compared to the $TiO_2$. Additionally, the color bars indicate the numerical value of energy being boosted into the cavity by ~13 and ~12 folds, respectively.

## S7. Reflection and electric field distributions for the $TiO_2$ cavity thickness of 200 nm

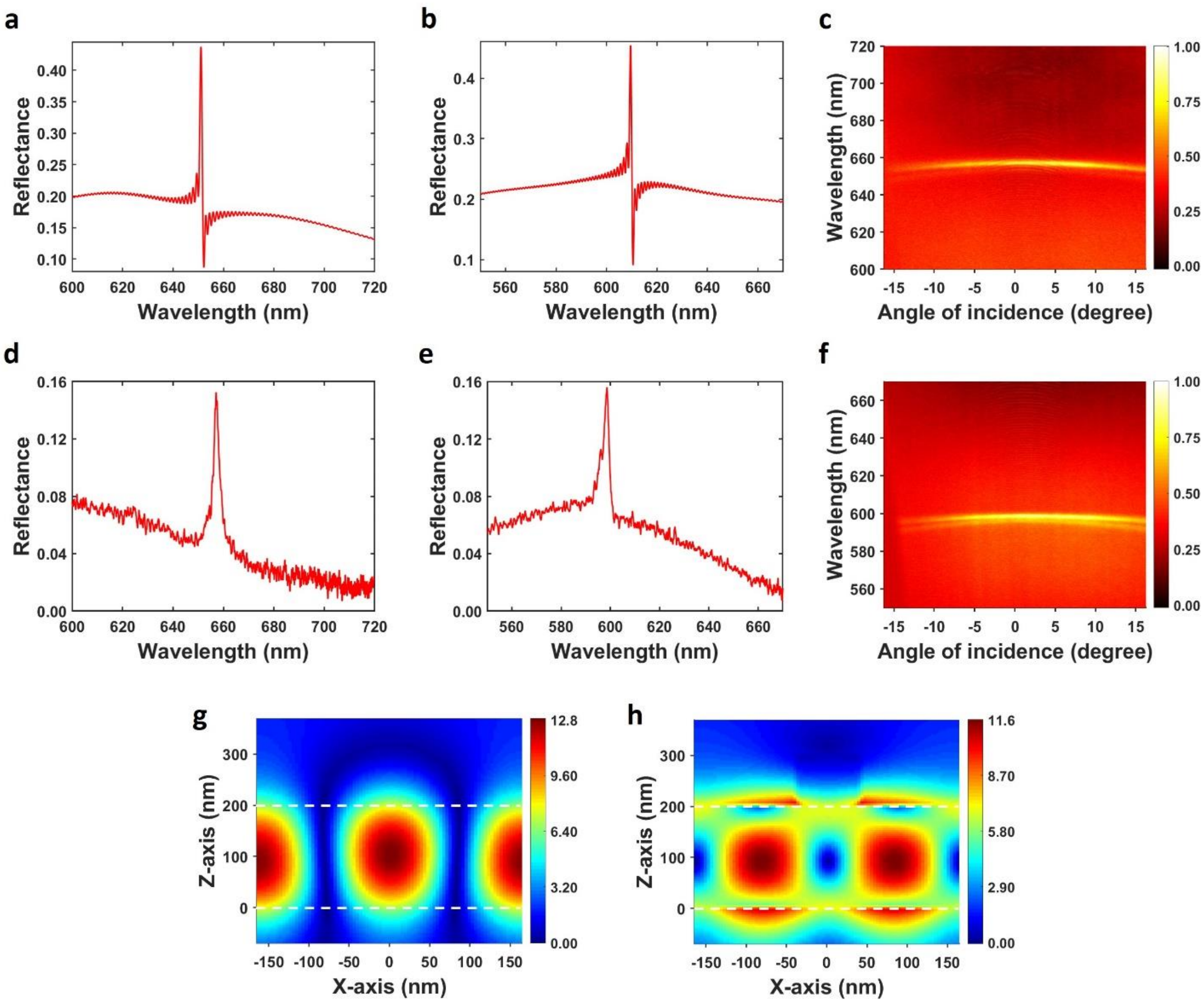

**Figure S7.** Reflectance spectra, angle-resolved reflectance, and electric field maps from the vertical cross-section where the cavity layer is $TiO_2$ with a thickness of 200 nm. (a, b, d, e) Numerical (a, b) and experimental (d, e) reflectance spectra for the electric field polarization of the applied source along the y-axis (a, d) and along the x-axis (b, e), where the pronounced Fano dips at 651.4 (a), and 609.9 nm (b), respectively, and the Lorentzian shape resonances at 656.0 (d), and 598.0 nm (d), respectively. (c, f) Normalized experimental angle-resolved reflectance for the electric field polarization of the source along the y-axis (c) and x-axis (f), demonstrating the wide-angle stability of resonances. (g, h) Electric field map from the vertical cross-section of the metasurface at the resonance point of 651.4 (a) and 609.9 nm (b), respectively, denoting the light energy completely accumulated into the cavity region (g) and both inside and outside of the cavity (h) [Here dashed lines denote the cavity region].

The numerical and experimental reflectance, along with the experimental angle-resolved reflectance and numerical cross-sectional electric field maps for a cavity thickness of 200 nm, which is composed of $TiO_2$ only, are presented in Figure S7. The profound Fano dips are evident in Figures S7(a) and S7(b), with Q-factors of 1,002 and 1,000, respectively. However, the experimental reflectance given in Figures S7(d) and S7(e) does not show the Fano dips, unlike other structures. Similarly, the experimental angle-resolved reflectance does not exhibit the sharp changes in reflectance observed in Figures S7(c) and S7(f). The numerical electric field maps in Figures S7(g) and S7(h) demonstrate light energy enhancements of ~13 and ~12 folds, respectively, within the cavity region.

## S8. Reflection and electric field distributions for the $TiO_2$ cavity thickness of 270 nm

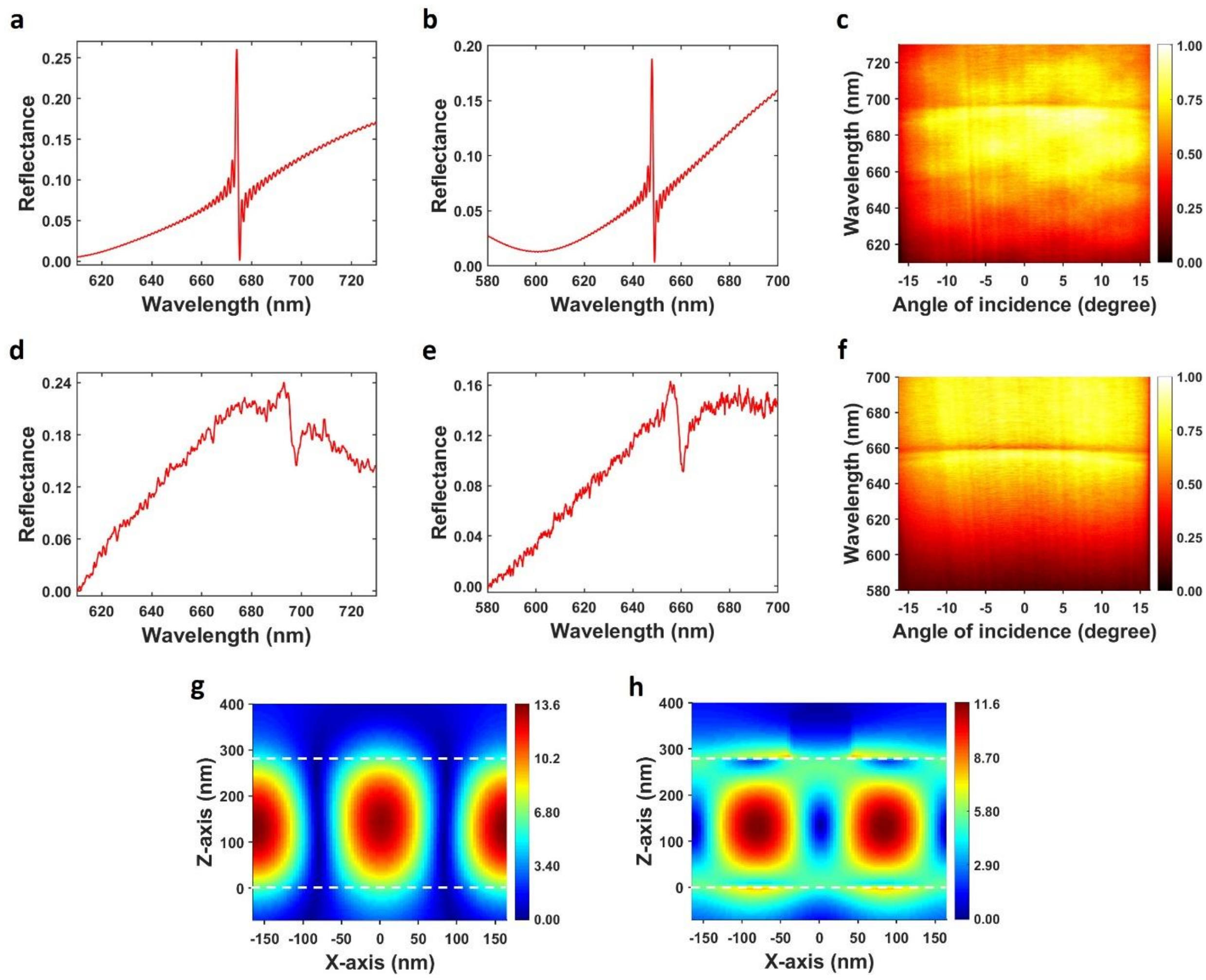

**Figure S8.** Reflectance spectra, angle-resolved reflectance, and electric field maps from the vertical cross-section where the cavity layer is $TiO_2$ with a thickness of 270 nm. (a, b, d, e) Numerical (a, b) and experimental (d, e) reflectance spectra for the electric field polarization of the applied source along the y-axis (a, d) and along the x-axis (b, e), where the pronounced Fano dips at 674.6 (a), 648.4 (b), 694.0 (d), and 660.0 nm (d), respectively. (c, f) Normalized experimental angle-resolved reflectance for the electric field polarization of the source along the y-axis (c) and x-axis (f), demonstrating the wide-angle stability of Fano resonances. (g, h) Electric field maps from the vertical cross-section of the metasurface at the resonance dip point of 674.6 (a) and 648.4 nm (b), respectively, denoting the light energy completely trapped into the cavity region [Here dashed lines denote the cavity region].

The final tuning example of our research work is a design where the cavity material is composed entirely of $TiO_2$, with a thickness of 270 nm. This design exhibits Fano dips at 674.6 and 648.4 nm, respectively, with high Q-factors of 1,007 and 954, as shown in Figures S8(a) and S8(b). The corresponding experimental reflectances are presented in Figures S8(d) and S8(e), with relatively low Q-factors of 217 and 192, respectively. However, there are significant differences in spectral positions between the numerical and experimental results, which are around 20 nm between Figures S8(a) and S8(d) and 12 nm between Figures S8(b) and S8(e). In addition, the intensity difference between the dip and the peak of the Fano line curve is not very sharp compared to other tuning examples (Figures S6(c) and S6(f)). The electric field cross-sectional maps display a similar field pattern to those shown in Figure S7, with enhancement of ~14 (Figure S8(g)) and ~12 folds (Figure S8(h)), respectively.

## S9. Reflections after the changing of the grating width

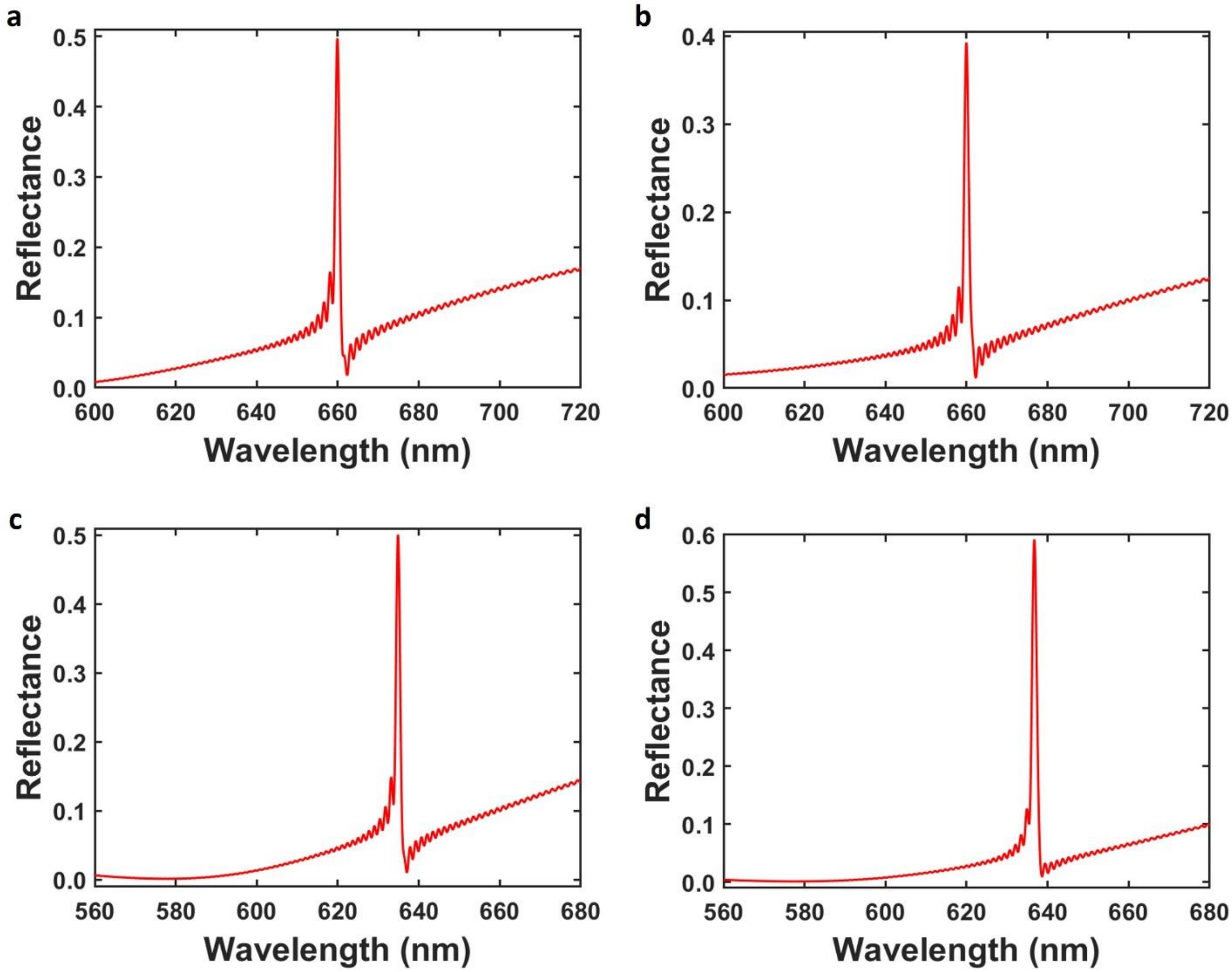

**Figure S9.** Numerical reflectance from the tuning of the grating width. (a, b) Reflectance spectra from the grating widths of 140 and 200 nm, respectively, while the polarization of the applied source is along the y-axis. (c, d) Corresponding reflectance spectra for the polarization direction along the x-axis.

The effect of grating width variation is demonstrated in Figure 5, where pronounced asymmetric Fano line shapes appear between the widths of 40 and 120 nm. Additionally, we observed that the Fano shapes transition to Lorentzian lines gradually as the grating width increases. We numerically performed the reflectance calculations for widths of 140 and 200 nm to support this claim. From Figure S9, it is confirmed that the Fano shapes are gradually transitioning into Lorentzian shapes, with the Fano dips disappearing as the grating width increases.